\documentclass[12pt]{iopart}
\usepackage{latexsym}
\usepackage{graphics}
\usepackage{graphicx}
\usepackage{amssymb}
\usepackage[latin1]{inputenc}

\newcommand{\mgb}{MgB$_2$}
\newcommand{\hcu}{$H_{c2}$}
\newcommand{\hcp}{$H_{c2}^{\pi}$}
\newcommand{\tc}{$T_{c}$}
\begin{document}

\title[Magnetic hysteresis in the mw surface resistance of MgB$_2$ at fields lower than $H_{c2}^{\pi}$]{Field-induced suppression of the $\pi$-band superconductivity and magnetic hysteresis in the microwave surface resistance of MgB$_2$ at temperatures near $T_c$}

\author{M. Bonura, A. Agliolo Gallitto and M. Li Vigni}
\address{CNISM and Dipartimento di Scienze Fisiche e Astronomiche, Università di Palermo, Via Archirafi 36, I-90123 Palermo (Italy)}

\author{G. A. Ummarino}
\address{CNISM and Dipartimento di Fisica, Politecnico di Torino, Corso Duca degli Abruzzi 24, I-10129 Torino, Italy}

\begin{abstract}
We report on the magnetic-field-induced variations of the microwave surface resistance, $R_s$, in a polycrystalline  \mgb\ sample, at different values of temperature. We have detected a magnetic hysteresis in $R_s$, which exhibits an unexpected plateau on decreasing the DC magnetic field below a certain value. In particular, at temperatures near $T_c$ the hysteresis manifests itself only through the presence of the plateau. Although we do not quantitatively justify the anomalous shape of the magnetic hysteresis, we show that the results obtained in the reversible region of the $R_s(H)$ curve can be quite well accounted for by supposing that, in this range of magnetic field, the $\pi$ gap is almost suppressed by the applied field and, consequently, all the $\pi$-band charge carriers are quasiparticles. On this hypothesis, we have calculated $R_s(H)$ supposing that fluxons assume a conventional (single core) structure and the flux dynamics can be described in the framework of conventional models. From the fitting of the experimental results, we determine the values of $H_{c2}^{\pi}(T)$ at temperatures near $T_c$. In our opinion, the most important result of our investigation is that, at least  at temperatures near $T_c$,  the value of the applied field that separates the reversible and irreversible regions of the $R_s(H)$ curves is just $H_{c2}^{\pi}(T)$; a qualitative discussion of the possible reason of this finding is given.
\end{abstract}




\maketitle

\section{Introduction}
A suitable method to investigate fluxon dynamics in type-II superconductors consists in measuring the magnetic-field-induced variations of the microwave (mw) surface resistance, $R_s$~\cite{golo,GOLOSOVSKY,noistatocritico,noiisteresi}. Indeed, the field dependence of $R_s$ in superconductors in the mixed state is determined by the presence of fluxons, which bring along normal fluid in their cores, as well as the fluxon motion. The experimental results are generally discussed in the framework of the two-fluid model, including the field dependence of the quasiparticle density and the effects of the fluxon motion~\cite{CC,BRANDT,dulcicvecchio}.

Studies reported in the literature on the field-induced variations of $R_s$ in \mgb\ have highlighted several anomalies, especially at applied magnetic fields much lower than the upper critical field, among which unusually enhanced field-induced mw losses~\cite{shibata,nova,dulcic,noi-irr} and a magnetic hysteresis of unconventional shape~\cite{noi-irr,isteresiMgB2,EUCAS2007}. It has been suggested that these anomalies are strictly related to the peculiarities of the fluxon lattice in \mgb. On the other hand, it is by now accepted that fluxons in \mgb\ have a composite structure, being constituted by two concentric cores, one of radius $\xi_{\sigma}$ (small core), associated with the $\sigma$ gap ($\Delta_{\sigma}$), and the other of radius $\xi_{\pi}$ (giant core), associated with the $\pi$ gap ($\Delta_{\pi}$)~\cite{eskil,nakai,koshelev}. Moreover, the gaps depend very differently on the magnetic field: while the $\sigma$ gap closes at the macroscopic upper critical field, \hcu, the $\pi$ gap is almost closed at magnetic-field values much lower than $H_{c2}$. This crossover field, at which the contribution to the superconductivity due to the $\pi$ superfluid is negligible, has been highlighted in several experiments~\cite{eskil,bouquet-H,Daghero2,Samuely} and is indicated as $H_{c2}^{\pi}$. Because of the different magnetic-field dependence of the two gaps, on varying the field, the structure of the vortex lattice evolves in an unusual way. At low magnetic fields, quasiparticles from $\pi$ and $\sigma$ bands are trapped within the vortex core, even if on different spatial scales. On increasing the field, though $\sigma$-band quasiparticles remain localized in the small core, giant cores start to overlap because of the field-induced suppression of $\Delta_{\pi}$; eventually, when $H_{c2}^{\pi}$ is reached, $\pi$-band quasiparticles are spread over the whole sample~\cite{eskil}. On further increasing the field, the $\pi$-quasiparticle density remains almost unchanged, while the $\sigma$-quasiparticle density continues to increase up to the macroscopic $H_{c2}$~\cite{diffusivita}. Only at applied fields higher than $H_{c2}^{\pi}$, fluxons assume a more conventional shape (single core), but they are surrounded by both the condensed fluid of the $\sigma$ band and the normal fluid of the $\pi$ band. This field-induced evolution of the vortex structure is expected to affect both the vortex-vortex and the vortex-pinning interactions, making the standard models inadequate to describe the fluxon dynamics in \mgb\ in a wide range of magnetic fields. Sarti et \emph{al}.~\cite{Sarti}, investigating the mw surface impedance of \mgb\ films, have shown that at low fields, when the contribution of the $\pi$-band superfluid cannot be neglected, the magnetic-field dependence of the real and imaginary components of the surface impedance exhibit several anomalies. However, they have shown that for fields higher than a threshold value, lower than \hcu($T$), the experimental results can be justified in the framework of a generalized two-fluid model in which, under the hypothesis that the $\pi$-quasiparticle density has reached the saturation value, the contribution of quasiparticles coming from the $\pi$ band is kept constant.

In this paper, we report on the magnetic-field dependence of the mw surface resistance of a polycrystalline sample of \mgb. The field-induced variations of $R_s$ have been investigated by the cavity perturbation technique, using a copper cavity, at increasing and decreasing the DC magnetic field, $H_0$. The $R_s(H_0)$ curves exhibit several anomalies that cannot be justified in the framework of the standard theories for the fluxon dynamics, among which a magnetic hysteresis of unconventional shape: in the decreasing-field branch of the $R_s(H_0)$ curve, we have detected an unexpected plateau extending from a certain value of the magnetic field down to zero. The hysteresis is detectable up to temperatures close to $T_c$ ($T/T_c\approx 0.95)$; however, the extension of the plateau depends on $T$ and, in particular, for $T\geq 0.77~T_c$ the hysteresis manifests itself only through the presence of the plateau. In a previous paper~\cite{noi-irr}, we have extensively discussed the anomalies of the $R_s(H_0)$ curves in \mgb\ obtained at low temperatures and we have ascribed them to the unusual vortex structure in this compound. In this paper, we devote attention to the range of fields at which the superconductivity coming from the $\pi$ band is almost suppressed. In this region ($H_0 \geq H_{c2}^{\pi}$), we expect that the flux lines assume a conventional single-core structure and that all the charge carriers coming from the $\pi$ band are quasiparticles. On this hypothesis, we have modified the expression of the complex penetration depth of the mw field, considering that the contribution of the  $\pi$ band to the field-induced energy losses is merely due to the presence of the $\pi$ quasiparticles and that of the $\sigma$ band is due to both the $\sigma$ quasiparticles and the fluxon motion. We will show that the experimental results can be quite well justified, in the framework of the proposed model, for fields greater than a threshold field, which we infer to be $H_{c2}^{\pi}$. The fitting of the experimental data allows determining with good accuracy the temperature dependence of $H_{c2}^{\pi}$ at temperatures near \tc, where other techniques fail in giving accurate results. Furthermore, our results show that, at least at temperatures near $T_c$, the values of $H_{c2}^{\pi}(T)$ coincide with the values of the applied magnetic field below which the decreasing-field branch of the $R_s(H_0)$ curve exhibits the plateau.

\section{Experimental apparatus and sample}\label{sec:samples}

\begin{figure}[h]
\centering
\includegraphics[width=7cm]{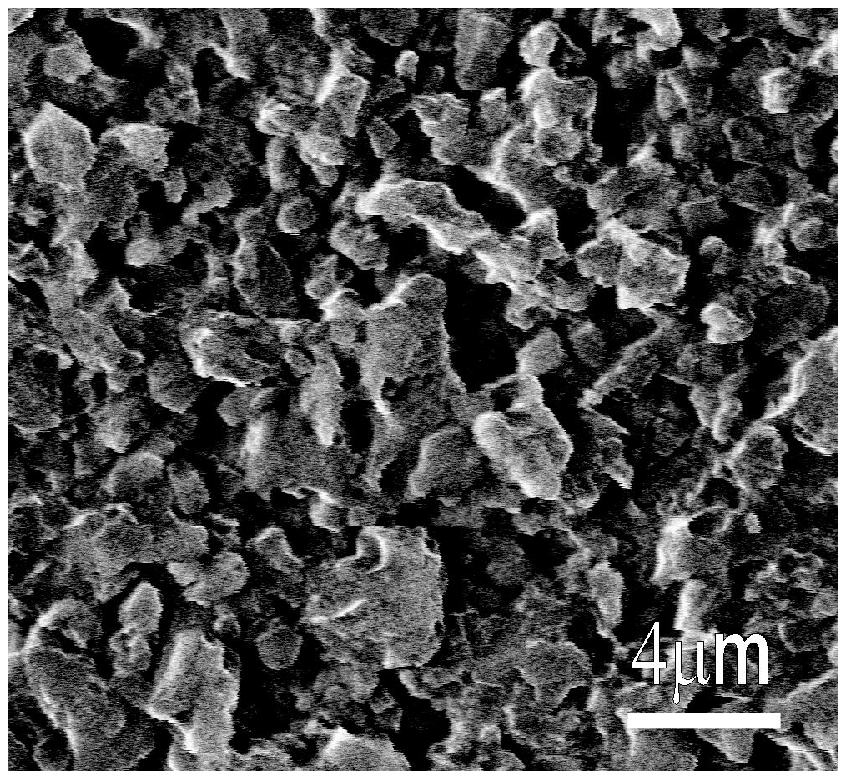}
\caption{SEM micrograph of the sample surface.}
\label{immagine}
\end{figure}

The field-induced variations of the mw surface resistance have been investigated in a bulk sample of $\mathrm{Mg}^{10} \mathrm{B}_2$ prepared at the INFM-LAMIA/CNR laboratory in Genova using the so-called one-step method~\cite{palenzona}. The sample has parallelepiped shape with approximate dimensions $2 \times 3 \times 0.5$ mm$^3$; it
undergoes a sharp superconducting transition with onset $T_c \approx 38.9$~K and $\Delta T_c \approx 0.3$~K. Figure~\ref{immagine} shows the SEM micrograph of the characteristic morphology of the sample surface; it highlights pores of the order of $\sim 1~\mu$m.

The mw surface resistance has been measured by the cavity-perturbation technique. A copper cavity, of cylindrical shape with golden-plated walls, is tuned in the $\mathrm{TE}_{011}$ mode resonating at $\omega/2\pi \approx 9.6$~GHz ($Q \approx 40000$ at LHe temperature). The sample is located in the center of the cavity, by a sapphire rod, where the mw magnetic field is maximum. The cavity is placed between the poles of an electromagnet which generates DC magnetic fields up to 1~T. Two additional coils, independently fed, allow compensating the residual field and working at low magnetic fields. A LHe cryostat and a temperature controller allow working either at fixed temperatures or at temperature varying with a constant rate.

\begin{figure}[h]
\centering
\includegraphics[width=8cm]{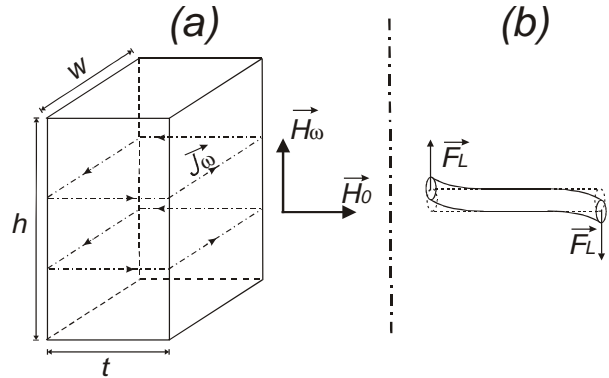}
\caption{(a) Field and current geometry at the sample surface; $w \approx 2~\mathrm{mm}, t \approx 0.5~\mathrm{mm}, h \approx 3~\mathrm{mm}$. (b)
Schematic representation of the motion of a flux line.}
\label{sample}
\end{figure}

The sample and the field geometries are schematically shown in Figure~\ref{sample}a; the DC magnetic field, $\vec{H}_{0}$, is perpendicular to the mw magnetic field, $\vec{H}_{\omega}$. When the sample is in the mixed state, the induced mw current causes a tilt motion of the vortex lattice~\cite{BRANDT}; Figure~\ref{sample}b schematically shows the motion of a flux line.

The surface resistance of the sample is given by
\begin{equation*}
    R_s= \Gamma \left(\frac{1}{Q_L} - \frac{1}{Q_U}\right)\,,
\end{equation*}
where $Q_L$ is the quality factor of the cavity loaded with the sample, $Q_U$ that of the empty cavity and $\Gamma$ the geometry factor of the sample.\\ The quality factor of the cavity has been measured by an hp-8719D Network Analyzer. All the measurements have been performed at very low input power; the estimated amplitude of the mw magnetic field in the region in which the sample is located is of the order of $0.1~\mu$T.

\section{Experimental results}\label{experimental}
The field-induced variations of $R_s$ have been investigated for different values of the temperature. For each measurement, the sample was zero-field cooled (ZFC) down to the desired temperature; the DC magnetic field was increased up to a certain value and, successively, decreased down to zero.

\begin{figure}[ht]
\begin{center}
\includegraphics[width=8cm]{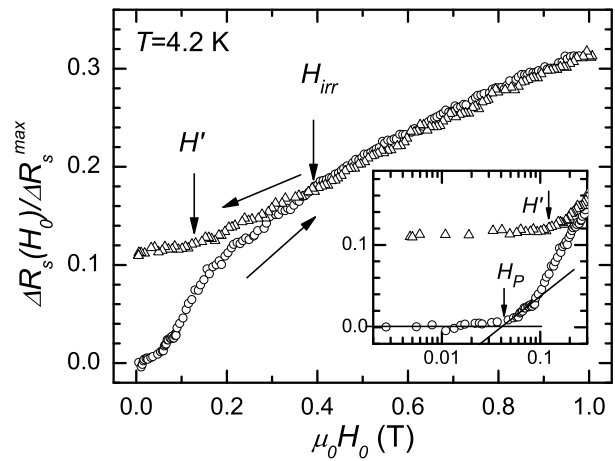}
\end{center}
\caption{\label{Rs(T4)}Field-induced variations of $R_s$ at $T=4.2$ K obtained by sweeping $H_0$ from 0 to 1 T (circles) and back (triangles). The inset shows the results in a logarithmic scale to better identify $H_p$. }
\end{figure}

Figure~\ref{Rs(T4)} shows the field-induced variations of $R_s$ at $T=4.2$~K, obtained by sweeping $H_0$ from zero to 1~T and back. In the figure, $\Delta R_s(H_0)\equiv R_s(H_0,T)-R_{res}$ and $\Delta R_s^{max}\equiv R_{n}-R_{res}$, where $R_{res}$ is the residual mw surface resistance at $T=2.5$~K and $H_{0}=0$, and $R_n$ is the normal-state surface resistance at $T=T_c$. The inset shows the results of the increasing-field branch in a logarithmic scale, which allows identifying the value of the applied magnetic field at which $R_s$ deviates from its zero-field value; this value should be the first-penetration field of vortices, $H_p$.  The decreasing-field branch is characterized by two characteristic fields: i) $H_{irr} $, which indicates the value of the applied field that separates the reversible and irreversible part of the $R_s(H_0)$ curve; ii) $H ^{\prime}$, which identifies the beginning of a plateau in the curve. Magnetic hysteresis in $R_s$ is expected as a consequence of the critical state of the vortex lattice; it has been detected in other superconductors~\cite{noiisteresi,Ji,sridhar} and has been ascribed to the different $B$ value at increasing and decreasing fields. However, as we will discuss in more detail in the next section, the presence of the plateau is puzzling because it should indicate that the trapped flux does not change anymore on decreasing the field below $H ^{\prime}$.

\begin{figure}[h!]
\begin{center}
\includegraphics[width=8cm]{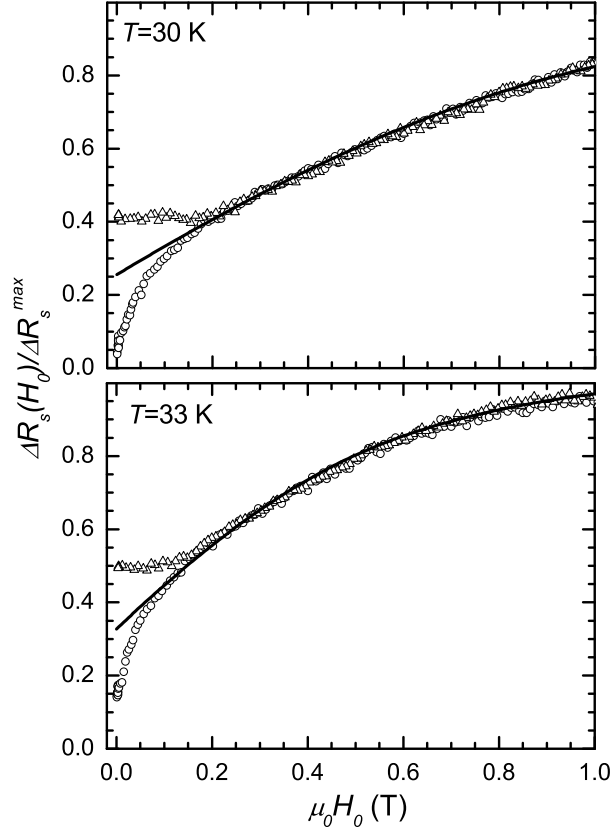}
\end{center}
\caption{\label{T30-33}Field-induced variations of $R_s$ obtained by sweeping $H_0$ from 0 to 1~T (circles) and back (triangles). The continuous lines are the best-fit curves obtained as described in Sec.~\ref{analysis}.}
\end{figure}

\begin{figure}[ht]
\begin{center}
\includegraphics[width=8cm]{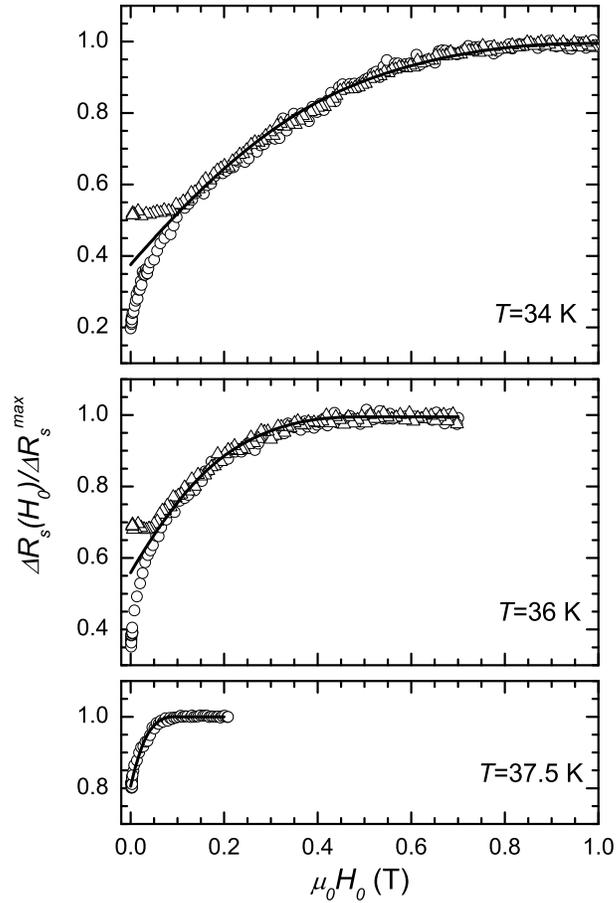}
\end{center}
\caption{\label{T>33}Field-induced variations of $R_s$ obtained by sweeping $H_0$ from 0 to $H_{max}$ (circles) and back (triangles), with $H_{max} > H_{c2}(T)$. The continuous lines are the best-fit curves obtained as described in Sec.~\ref{analysis}.}
\end{figure}

For $T \lesssim 0.77~T_c$, the decreasing-field branch of the $R_s(H_0)$ curve shows peculiarities similar to those of Figure~\ref{Rs(T4)}, with the two characteristic fields, though they decrease on increasing $T$. For $T \gtrsim 0.77~T_c$,  $H_{irr}(T)$ coincides with $H^{\prime}(T)$, i.e. the hysteresis manifests itself only through the presence of the plateau below $H^{\prime}(T)$; Figures~\ref{T30-33} and \ref{T>33} show the field-induced variations of $R_s$, just in this range of temperatures. In all the figures, $\Delta R_s(H_0)\equiv R_s(H_0,T)-R_{res}$ and $\Delta R_s^{max}\equiv R_{n}-R_{res}$. The continuous lines reported in the figures are the best-fit curves obtained as described in Section~\ref{analysis}.

\begin{figure}[h!]
\begin{center}
\includegraphics[width=8cm]{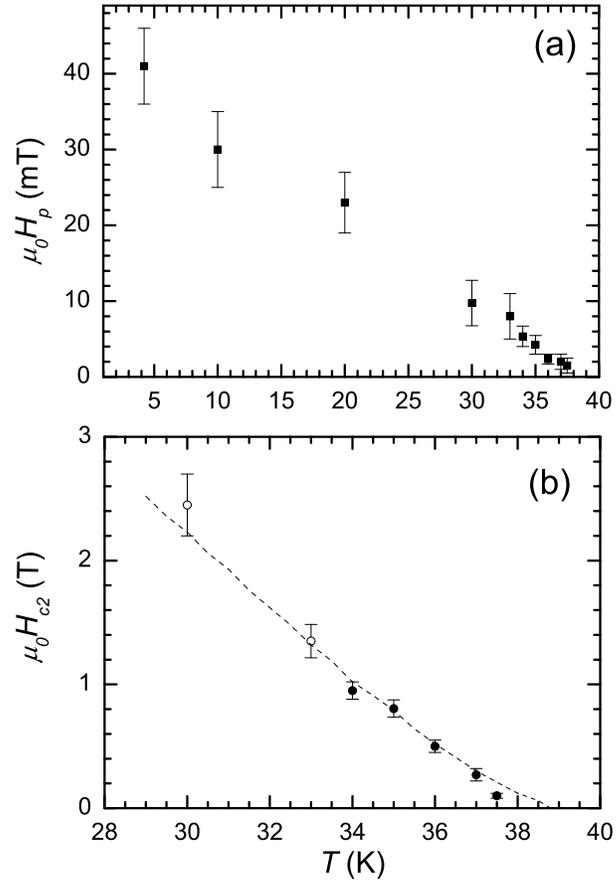}
\end{center}
\caption{\label{Hc2(T)}Temperature dependence of the first-penetration field (a) and the upper
critical field (b), deduced from the isothermal $R_s(H_0)$ curves. Full circles have been directly deduced from the experimental data; open circles have been determined by fitting the data. In any case, they identify $H_{c2}^{\perp c}(T)$. Dashed line is the curve calculated by the generalized two-band Eliashberg theory, as reported in the Appendix.}
\end{figure}

The zero-field value of the increasing-field branch of the $R_s(H_0)$ curve is determined by the contribution of quasiparticles due to thermal breaking of Cooper pairs. At $H_0=H_p$, vortices start to penetrate the sample giving rise to an increase of $R_s$. When the applied magnetic field reaches the value of the upper critical field, $R_s$ assumes its normal-state value, $R_n$. Figure~\ref{Hc2(T)} shows the temperature dependence of $H_p$ and $H_{c2}$ deduced from the isothermal $R_s(H_0)$ curves. The values indicated in panel (b) as full circles have been deduced measuring the magnetic field at which $R_s$ reaches $R_n$, those indicated as open circles have been obtained by fitting the experimental data as discussed in Sec.~\ref{analysis}. In any case, since our sample is a polycrystal the upper critical field so determined coincides with $H_{c2}^{\perp c}$. Dashed line is the curve calculated by the generalized two-band Eliashberg theory, as reported in the Appendix.

From Figure~\ref{Hc2(T)}a, one can see that $H_p$ exhibits a linear temperature dependence down to low
temperatures, consistently with results in \mgb\ samples reported by different authors~\cite{Lyard,Li,sharoni}. As it is known, the first-penetration field may differ from $H_{c1}$ if demagnetization and/or surface-barrier effects come into play; however, our results are consistent with the lower critical field for \mgb\ bulk samples reported in the literature~\cite{sharoni}.

By looking at Figures~\ref{T30-33} and~\ref{T>33}, one can see that the hysteretic behaviour of the $R_s(H_0)$ curve is detectable up to temperatures very close to $T_c$, although restricted in a narrow field range; only at $T\geq37.5$~K, the $R_s(H_0)$ curve is reversible in the whole range of fields investigated. This result suggests that pinning is effective up to temperatures close to $T_c$.

\section{Discussion}
Models for the electrodynamics of superconductors in the mixed state have been proposed by several authors~\cite{noistatocritico,noiisteresi,CC,BRANDT,dulcicvecchio}, who discuss the field-induced mw losses in different approximations. Coffey and Clem (CC) have elaborated a comprehensive theory, in the framework of the two-fluid model of  superconductivity, in which, besides the effects of the fluxons motion, the field dependence of the densities of the normal and condensed fluids are considered~\cite{CC}. The CC theory has been developed under the basic assumption that vortices generate a magnetic induction field, $B$, uniform in the sample. On this hypothesis, the mw surface impedance in the linear approximation ($H_{\omega} \ll\ H_0$) is given by
\begin{equation}\label{Rs}
    R_s=-\mu_{0}\omega ~\mathrm{Im}[{\widetilde{\lambda}(\omega,B,T)}]\, ,
\end{equation}
with
\begin{equation}\label{lambdat}
    \widetilde{\lambda}(\omega,B,T)=\sqrt{\frac{\lambda^{2}(B,T)+
    (i/2)\widetilde{\delta}_{v}^{2}(\omega,B,T)}
    {1-2i\lambda^{2}(B,T)/\delta _{nf}^{2}(\omega,B,T)}}\, .
\end{equation}
In Eq.~(\ref{lambdat}), $\widetilde{\delta} _{v}(\omega, B, T)$ is the effective complex skin depth arising from the vortex motion~\cite{CC,noiBKBO}, while the temperature and field dependencies of the two fluids are taken into account through $\lambda(B,T)$ and $\delta_{nf}(\omega,B,T)$, given by
\begin{equation}\label{8-lamda0}
\lambda(B,T) = \frac{\lambda_0}{\sqrt{n_s(B,T)}},
\end{equation}
\begin{equation}\label{8-delta0}
\delta_{nf}(\omega,B,T) =
\frac{\delta_0}{\sqrt{n_n(B,T)}}=\sqrt{\frac{2}{\mu_0 \omega
\sigma_n(B,T)}}\, ,
\end{equation}
where $\lambda _0$ is the London penetration depth at $T=0$; $\delta _0$ is the normal skin depth at $T=T_c$; $n_s(B,T)$ and $n_n(B,T)$ are the condensed- and normal-fluid fractions, respectively; $\sigma_n(B,T)$ is the normal-fluid conductivity in the superconducting state.\\

Equations~(\ref{Rs}--\ref{8-delta0}) have been obtained under the hypothesis that $B$ is uniform inside the sample. When the fluxon lattice is in the critical state, the assumption of uniform $B$ is no longer valid; as a consequence, the hysteresis in the $R_s(H_0)$ curve cannot be justified by Eqs.~(1--4). In our field geometry (see Fig.~\ref{sample}a), the effects of the non-uniform $B$ distribution on $R_s$ are particularly enhanced because in the two sample surfaces normal to the external magnetic field the
mw current and fields penetrate along the fluxon axis and, consequently, the mw losses involve the whole vortex lattice. However, in this case, one can easily take into account the non-uniform $B$ distribution by calculating a proper averaged value of $R_s$ over the whole sample as follows~\cite{noistatocritico,noiisteresi}
\begin{equation}\label{RsMED}
    R_{s}= \frac{1}{S}\int_\Sigma R_s(|B(\vec{r})|)\,
    dS\,,
\end{equation}
where $\Sigma$ is the sample surface, $S$ is its area and $\vec{r}$ identifies the surface element.\\
It is worth noting that, in order to use Eq.~(\ref{RsMED}) for taking into due account the critical-state effects, it is essential to know the $B$ profile inside the sample, determined by the field dependence of the critical current density.

Recently, using this method, we have investigated the effects of the critical state on the field-induced variation of $R_s$, at increasing and decreasing fields~\cite{noistatocritico,noiisteresi}. We have shown that the parameter that mainly determines the peculiarities of the $R_s(H_0)$ curve is the full penetration field, $H^*$. In particular, $H^*$ strongly affects the amplitude of the hysteresis loop and the shape of the increasing-field branch of the $R_s(H_0)$ curve. However, independently of the $H^*$ value, the shape of the decreasing-field branch, being strictly related to the shape of the magnetization curve, should exhibit a negative concavity, with a monotonic reduction of $R_s$ from $H_0=H_{max}$ down to $H_{c1}$.

The $R_s(H_0)$ curves we have obtained cannot be justified in the framework of the model above mentioned, for several reasons. From Figure~\ref{Rs(T4)}, one can see that the application of a magnetic field of $\approx 1$~T, which is of the order of $H_{c2}/10$, causes a $R_s$ variation larger than $30\%$ of the maximum variation. These field-induced variations of $R_s$ are much greater than those expected from the models reported in the literature~\cite{noistatocritico,noiisteresi,CC,BRANDT,dulcicvecchio} and detected in other superconductors~\cite{noiisteresi,noiBKBO,TALVA}. Another anomaly concerns the shape of the magnetic hysteresis and, in particular, the presence of the plateau in the decreasing-field branch. As above mentioned, the decreasing-field branch of the $R_s(H_0)$ curve is expected to exhibit a monotonic reduction of $R_s$ down to $H_{c1}$. The presence of the plateau is puzzling; indeed, it would suggest that the trapped flux does not change anymore on decreasing the field below $H^{\prime}$, although this value is much larger than the first penetration field.

We would like to remark that results similar to those reported in this paper have been observed in other \mgb\ bulk samples, produced by different methods~\cite{noi-irr,EUCAS2007}. We have extensively discussed the anomalies of the $R_s(H_0)$ curve in Refs.~\cite{noi-irr,EUCAS2007}, and we have suggested that they are ascribable to the unusual properties of the fluxon lattice in two-gap \mgb. This hypothesis has been corroborated by the fact that the results obtained in a strongly irradiated \mgb\ sample, in which the two gaps merged into a single value, have been quite well accounted for in the framework of the model above discussed, using Eqs.~(\ref{Rs}--\ref{RsMED})~\cite{noi-irr,noi-depin}. In two-gap \mgb\ samples, according to Ref.~\cite{shibata}, the unusually enhanced field-induced variations of $R_s$ at applied fields much lower than $H_{c2}$ have been qualitatively ascribed to the presence and motion of the giant cores due to the $\pi$-band quasiparticles. On the contrary, the origin of the anomalous shape of the $R_s(H_0)$ curve is so far not understood. Only in finely powdered \mgb\ samples the hysteresis assumes a more conventional shape (without plateau), suggesting that the presence of the plateau is related to a bulk process~\cite{EUCAS2007}.

By looking at Figures~\ref{T30-33} and \ref{T>33}, one can see that the hysteretic behaviour of the $R_s(H_0)$ curve is detectable up to $T \approx 0.95~T_c$, suggesting that pinning is effective up to temperatures very near $T_c$. However, on increasing the temperature the irreversibility is restricted in a narrow range of fields and, for $T \gtrsim 0.77~T_c$, it manifests itself by the mere presence of the plateau; in the following, we will focus on the analysis of the results obtained just in this temperature range.

\subsection{The model}\label{model}
The study of the field-induced variations of $R_s$ in \mgb\ is made particularly complex by the unconventional structure of vortices, which is expected to affect both the vortex-vortex and the vortex-defect interactions. However, for $H_0\geq H_{c2}^{\pi}$, vortices are expected to assume a "single-core" structure; indeed, giant cores overlap and $\sigma$ quasiparticles remain
localized within the small cores, which will overlap at $H_0 = H_{c2}$. Starting from this picture, we have modified the expression of the complex penetration depth of the mw field; we have considered that the contribution of the $\pi$ band to the field-induced energy losses is merely due to the presence of the $\pi$ quasiparticles and that of the $\sigma$ band is due to both the $\sigma$ quasiparticles and the fluxon motion. We can neglect the effects of the critical state if we look at the reversible region of the $R_s(H_0)$ curve.

Since in \mgb\ the superconductivity is due to the charge carriers coming from the two different bands, one can write: $n_s=n_s^{\pi}+n_s^{\sigma}$ and $n_n=n_n^{\pi}+n_n^{\sigma}$. For $H_{c2}^{\pi}\leq H_0\leq H_{c2}$, the superelectron fraction, $n_s$, reduces to $n_s^{\sigma}$ because $n_s^{\pi}=0$. So, $\lambda(B,T)$ is determined by the $\sigma$ band and Eq.~(\ref{8-lamda0}) can be rewritten as
\begin{equation}\label{8-lamda0-2}
\lambda(B,T) =
\frac{\lambda_0^{\sigma}}{\sqrt{n_s^{\sigma}(B,T)}}=\frac{\lambda_0^{\sigma}}{\sqrt{n_s^{\sigma}(0,T)[1-
B /B_{c2}(T)]}}\; .
\end{equation}

In \mgb, due to the presence of normal fluid coming from the two bands, the overall normal-state conductivity can be reasonably considered as the sum of the normal-state conductivities of the
$\pi$ and $\sigma$ bands: $\sigma_n=\sigma_n^{\pi}+\sigma_n^{\sigma}$.
Consequently, the normal skin depth can be written as
\begin{equation}\label{N-S-D}
\delta_0=\sqrt{\frac{2}{\mu_0 \omega
\sigma_n}}=\sqrt{\frac{2}{\mu_0 \omega
(\sigma_n^{\pi}+\sigma_n^{\sigma})}}\;.
\end{equation}
Sarti \textit{et al.}~\cite{Sarti} have suggested that, since the two bands interact very weakly each other \cite{Mazin:107002}, when superconductivity arising from the $\sigma$ band is suppressed the conductivity of the quasiparticle fraction can be written as
\begin{equation}\label{sigma-n}
\sigma_{nf}(B,T)=\sigma_n^{\pi}+\sigma_{nf}^{\sigma}(B,T)=\sigma_n^{\pi}+[1-n_s^{\sigma}(B,T)]\sigma_n^{\sigma}
\;.
\end{equation}
It follows that, in the range of fields considered, Eq.~(\ref{8-delta0}) can be modified as
\begin{equation}\label{8-delta0-2}
\delta_{nf}(\omega,B,T)=\frac{\delta_0}{\sqrt \frac{\sigma_{nf}(B,T)}{\sigma_n}}=
\frac{\delta_0}{\sqrt{1-n_s^{\sigma}(B,T)\frac{\sigma_n^{\sigma}}{\sigma_n}}}\,.
\end{equation}

In order to calculate the complex penetration depth of the mw field, besides $\lambda (B,T)$ and $\delta_{nf}(\omega,B,T)$, it is necessary to determine the effective complex penetration depth due to the vortex motion (see Eq.~\ref{lambdat}). $\widetilde{\delta}_{v}(\omega,B,T)$ depends on the relative magnitude of the viscous and restoring-pinning forces through the depinning frequency~\cite{gittle}. Since we are analyzing the results in a restrict range of temperature near $T_c$, and considering the values of the depinning frequency reported in the literature for \mgb~\cite{dulcic,silva}, it is reasonable to assume that the vortex motion at our working frequency is ruled by the viscous drag force, i.e. vortices move in the flux-flow regime. The hypothesis is strengthened by the consideration that for $H_0\geq H_{c2}^{\pi}$ vortices are surrounded by $\sigma$-band condensed fluid and $\pi$-band normal fluid; this should reduce the stabilization energy and, consequently, the pinning efficacy. In the flux-flow regime, the expression of $\widetilde{\delta}_{v}$ reduces to
\begin{equation}\label{deltav-ff}
\widetilde{\delta}_{v}(\omega,B,T)=\delta_0 \sqrt{B/B_{c2}(T)} \;.
\end{equation}

The equations reported up to now do not explicitly consider the anisotropy properties of \mgb. In order to take into account the anisotropy, one can assume that the polycrystalline sample is constituted by grains with the $c$-axis randomly oriented and suppose that at fixed temperatures the anisotropic Ginzburg-Landau theory can be used. By indicating with $\theta$ the angle between $\vec{H}_0$ and the $c$-axis of the generic crystallite, the upper critical field is given by
\begin{equation*}
    H_{c2}(\theta) = \frac{H_{c2}^{\perp
    c}}{\sqrt{\gamma^2 \cos^2(\theta) + \sin^2(\theta)}} \,,
\end{equation*}
where $\gamma$ is the anisotropy factor of the upper critical field.

Also the anisotropy of the penetration depth plays a role because it determines the sample surface layers in which the mw energy losses occur. Indicating with $\alpha$ the angle between $\vec{H}_{\omega}$ and $\hat{c}$ one can write
\begin{equation*}
    \lambda_0^{\sigma}(\alpha) = (\lambda_0^{\sigma})_{ab}\sqrt[4]{\cos^2(\alpha) + \gamma_{\lambda}^2\sin^2(\alpha)} \,,
\end{equation*}
where
\begin{equation*}
\gamma_{\lambda}=\frac{(\lambda_0^{\sigma})_c}{(\lambda_0^{\sigma})_{ab}}\,.
\end{equation*}

It is worth noting that, since in our field geometry $\vec{H}_0 \perp \vec{H}_{\omega}$, $\alpha \neq \theta$. Supposing $\vec{H}_0\equiv \hat{z}$ and $\vec{H}_{\omega}\equiv \hat{x}$, one can easily find $\alpha = \arccos (\sin \theta \cos \varphi)$, being $\theta$ and $\varphi$ the polar and azimuthal angles. The mw surface resistance of the generic crystallite depends on the angles $\theta$ and $\alpha$ between its $c$-axis and the directions of $\vec{H}_{0}$ and $\vec{H}_{\omega}$, respectively; the overall surface impedance of the sample can be obtained by integrating over the whole solid angle.

\subsection{Analysis of the results}\label{analysis}
The expected results of the normalized $R_s(H_0)$ curves depend on $(\lambda_0^{\sigma})_{ab} / \delta_0$,  $H_{c2}^{\perp c}$, $n_s^{\sigma}(0,T)$, $\sigma_n^{\sigma}/\sigma_n$, $\gamma$ and $\gamma_{\lambda}$.
Most of these quantities are known and/or deducible from the experimental data.

\begin{figure}[t]
\begin{center}
\includegraphics[width=8cm]{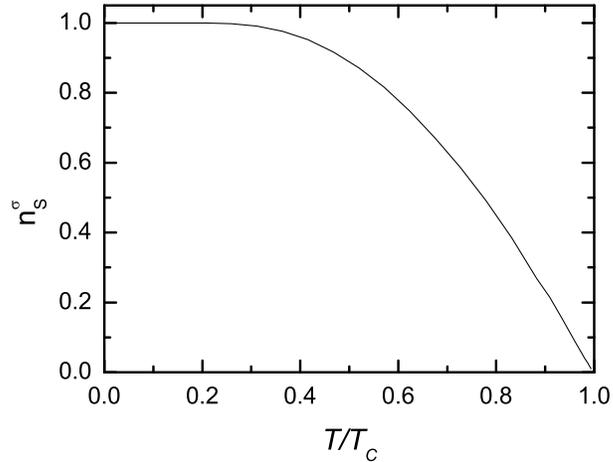}
\end{center}
\caption{\label{ns(T)}Temperature dependence of $n_s^{\sigma}$ at zero magnetic field, calculated by the generalized two-band Eliashberg theory, using the procedure reported in the Appendix.}
\end{figure}

The anisotropy of $\lambda_0^{\sigma}$ can be determined using the results of first-principles band-structure calculations~\cite{kortus}; in particular, from the values of the plasma frequencies reported by Brinkman \emph{et al.}~\cite{Brinkman}, one obtains $\gamma_{\lambda}=6.1$. The temperature dependence of $n_s^{\sigma}$ at zero magnetic field has been obtained by using the generalized two-band Eliashberg theory, which has already been used with success to study the MgB$_{2}$ system~\cite{carbinicol}. Figure~\ref{ns(T)} shows the calculated temperature dependence of $n_s^{\sigma}$; details of the calculation procedure are reported in the Appendix. In the analysis of our experimental results, we have used for $n_s^{\sigma}(0,T)$ the values reported in Figure~\ref{ns(T)} and for the anisotropy of $\lambda_0^{\sigma}$ the value $\gamma_{\lambda}=6.1$.

The anisotropy of the upper critical field, $\gamma$, has been determined both experimentally and theoretically~\cite{golubov,cubit,Lyard}; at temperatures near $T_c$, it should be in the range $2 \div 3$. In the temperature range $34 \div 37.5$~K, $H_{c2}^{\perp c}(T)$ has been deduced from the experimental data (full circles in Fig.~\ref{Hc2(T)}); for $T<34$~K it has to be taken as a parameter. Moreover, we have found that for $T\geq 33$~K the value of $\sigma_n^{\sigma}/\sigma_n$ affects very little the expected curves; this finding can be ascribed to the fact that at temperatures close to $T_c$, $\delta_{nf}(\omega, B, T) \approx \delta_0$.

In order to fit the results obtained in the temperature range $34 \div 37.5$~K, we have used for $H_{c2}^{\perp c}$ the values reported in Figure~\ref{Hc2(T)} as full circles, letting them vary within the experimental uncertainty; have let $\gamma$ vary in the range $2 \div 3$; have taken $(\lambda_0^{\sigma})_{ab} / \delta_0$ as free parameter. The best-fit curves are shown as continuous lines in Figure~\ref{T>33}; the values of the best-fit parameters are reported in Table~1.

For $T<34$~K, our apparatus does not allow measuring the upper critical field, which has to be taken as parameter as well; on the other hand, we can use for $(\lambda_0^{\sigma})_{ab} / \delta_0$ the value we have already determined by fitting the data at higher temperatures. At $T=30$~K, also the value of $\sigma_n^{\sigma}/\sigma_n$ affects in a not-negligible extent the expected curve; so, the parameters necessary to fit these data are  $\sigma_n^{\sigma}/\sigma_n$, $H_{c2}^{\perp c}$ and $\gamma$. The best-fit curves obtained at $T= 30$~K and $T=33$~K are shown in Figure~\ref{T30-33} and the parameters are listed in Table~1.
\begin{table}\centering
\begin{tabular}{ c c c c c }
  \hline   \hline
  $T$(K) & $(\lambda_0^{\sigma})_{ab}/ \delta_0$ & $\gamma$ & $\sigma_n^{\sigma}/\sigma_n$ & $H_{c2}^{\perp c}$~(T) \\
  \hline
  37.5 & $0.22\pm 0.02$ & $2.3\pm 0.2$ & --           & $0.10 \pm 0.01$ \\
  37 & $0.22\pm 0.02$   & $2.3\pm 0.2$ & --           & $0.35\pm 0.03$ \\
  36 & $0.22\pm 0.02$   & $2.4\pm 0.2$ & --           & $0.50\pm 0.05$ \\
  35 & $0.22\pm 0.02$   & $2.4\pm 0.2$ & --           & $0.90\pm 0.07$ \\
  34 & $0.22\pm 0.02$   &$2.5\pm 0.2$ & --           & $1.0 \pm 0.1$\\
  33 & 0.22             & $2.5\pm 0.2$ & --           & $1.35\pm 0.14$ \\
  30 & 0.22             & $2.7\pm 0.3$ & $0.10\pm 0.05$ & $2.4\pm 0.3$\\
  \hline   \hline
\end{tabular}
\caption{Values of the parameters that best fit the experimental data of Figs.~\ref{T30-33} and \ref{T>33}; the values reported without uncertainty have been imposed. For all the temperatures, we have used $\gamma_{\lambda}=6.1$ and $n_s^{\sigma}(0,T)$ of Fig.~7.}
\label{table-1}\end{table}

In the framework of the Eliashberg theory, we have calculated the temperature dependence of the upper critical field. We have used the linearized gap equations under magnetic field, for a superconductor in the clean limit~\cite{Sud}; details of the calculus are reported in the Appendix. In this calculation, the only input parameters are the Fermi velocities in the two bands. The calculated values of $H_{c2}^{\perp c}(T)$ are reported, as dashed line in Figure~\ref{Hc2(T)}; they have been obtained with $v^{\sigma}_{F ab}=4.4\times 10^{5}$~m/s and $v^{\pi}_{F ab}=20\times 10^{5}$~m/s. $v^{\sigma}_{F ab}$ is equal to that used in the band-theory calculus of Brinkman \emph{et al.}~\cite{Brinkman}; on the contrary, $v^{\pi}_{F ab}$ is larger than that reported by Brinkman \emph{et al.}, as obtained also by other authors~\cite{Sud}.

From the deduced value of $(\lambda_0^{\sigma})_{ab}/ \delta_0$, one can estimate $(\lambda_0^{\sigma})_{ab}$ if $\delta_0$ is known. The value of the normal-state mw surface resistance, we have measured at $T=T_c$, is $R_n \approx 40~\mathrm{m \Omega}$; from the estimated $R_n$ we deduce $\delta_0= 2 R_n / \mu_0 \mathrm{\omega} \approx 1~ \mu$m and, consequently, $(\lambda_0^{\sigma})_{ab} \approx 220$~nm. This value of $(\lambda_0^{\sigma})_{ab}$ is larger than the expected value calculated for a single crystal; we think that this is due to the porousness of the sample and/or the roughness of its surface (see Fig.~\ref{immagine}). Indeed, the surface roughness enlarges the effective area of the sample in which the mw energy losses occur, giving rise to an overestimated $R_n$ value, with consequent enlarged value of the deduced $\delta_0$ and  $(\lambda_0^{\sigma})_{ab}$.

\begin{figure}[htbp]
\begin{center}
\includegraphics[width=8cm]{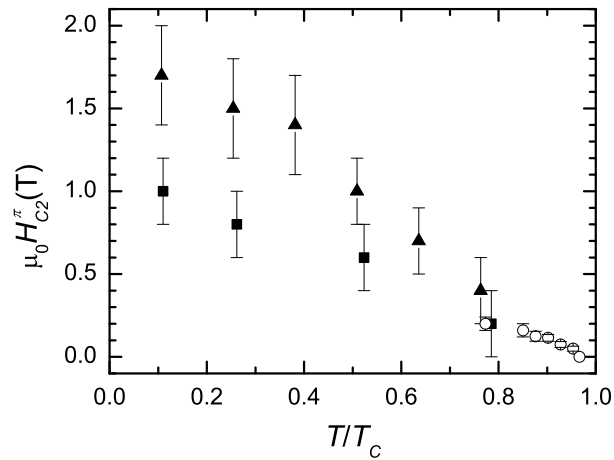}
\end{center}
\caption{Temperature dependence of $H_{c2}^{\pi}$. Open circles are the results we have obtained as
described in the text; full triangles and full squares are the results obtained by Daghero \textit{et al.} \cite{Daghero2} and Samuely \textit{et al.} \cite{Samuely}, respectively.}\label{Hc2Pi}
\end{figure}

Although the values of the best-fit parameters have been obtained by fitting the results for
$H_0>H^{\prime}$, the expected curves of Figures~\ref{T30-33} and~\ref{T>33} are reported in the whole range of fields investigated. In the framework of our model, the expected curves should properly describe the experimental results only for $H_0\geq H_{c2}^{\pi}$. At lower fields, it is expected that the theoretical curve overestimates $R_s(H_0)$ because our model assumes that all the $\pi$-band carriers are quasiparticles. Therefore, the $H_0$ value at which the theoretical curve begins to fit the experimental data can be taken as $H_{c2}^{\pi}(T)$. Using this criterion, we have deduced the temperature dependence of $H_{c2}^{\pi}$ in the range of temperatures investigated. The results are shown in Figure~\ref{Hc2Pi} as open circles; in the same figure, we report the values of $H_{c2}^{\pi}(T)$ obtained by Daghero \textit{et al.} (full triangles)~\cite{Daghero2} and Samuely \textit{et al.} (full squares)~\cite{Samuely} from point-contact spectroscopy.

The values we have obtained for $H_{c2}^{\pi}(T)$ match with those reported in Refs.~\cite{Daghero2} and \cite{Samuely} for lower temperatures; they give information on the temperature dependence of $H_{c2}^{\pi}$ in a range of temperatures close to $T_c$, where conventional techniques fail in
giving accurate results.

As one can see from Figures~\ref{T30-33} and \ref{T>33}, we have obtained good fits for $H_0>H^{\prime}(T)$; i.e. $H_{c2}^{\pi}(T)$ coincide with the values of $H_0$ at which the decreasing-field branch of the $R_s(H_0)$ curves deviates from the increasing-field one, within the experimental uncertainty. This finding suggests that the features of the $R_s(H_0)$ curves are strongly related to the suppression of the $\pi$ gap, due to the applied magnetic field. This is a curious result that needs further investigation; however, some considerations can be done.

As already mentioned, results very similar to those reported in this paper have been obtained in all the \mgb\ bulk samples we have investigated~\cite{noi-irr,isteresiMgB2,EUCAS2007}, except in a strongly irradiated \mgb\ bulk sample, in which the two gaps merged into a single value~\cite{noi-irr}. Moreover, the anomalous plateau in the decreasing-field branch of the $R_s(H_0)$ curve has not been observed in samples of \mgb\ consisting of fine powder~\cite{EUCAS2007}. In order to propose a possible explanation for the different results in bulk and powder samples, it is worth to recall that, in our field configuration ($\vec{H}_{0} \bot \vec{H}_{\omega}$), the mw current induces a tilt motion of the whole vortex lattice (see Fig.~\ref{sample}); so, our measurements allow detecting the response of the fluxon lattice also in the interior of the sample (far from the edges from which the DC magnetic field penetrates). On the other hand, in the framework of the critical state, the main property distinguishing the magnetic response of samples having small or large dimensions is the residual magnetic induction far from the edges, after the applied magnetic field has reached values larger than the full penetration field. We infer that the anomalous shape of the hysteresis in $R_s(H_0)$ detected in the \mgb\ bulk samples is related to a process occurring in the interior of the sample, probably due to the higher value of the local magnetic field.

The finding that the value of the applied field that separates the reversible and irreversible regions of the $R_s(H_0)$ curves coincides with $H_{c2}^{\pi}(T)$ can be justified by the consideration that for $H_0\geq H_{c2}^{\pi}$ vortices are surrounded by $\sigma$-band condensed fluid and $\pi$-band normal fluid; this should reduce the stabilization energy and, consequently, the pinning efficacy. The hysteretic behaviour at $H_0\leq H_{c2}^{\pi}$ could be justified by supposing that, after the sample has been exposed to fields higher than \hcp, on reducing $H_0$, the induction field inside the sample, far from the surface, is large enough to destroy the superconductivity in the $\pi$ band but, contemporaneously, the pinning is effective in the sample regions near the surface, where the local magnetic induction is smaller than $\mu_0$ \hcp. When the DC field is reduced below \hcp, the vortices near the surface experience the restoring pinning force which, hindering the free exit of fluxons, may keep the inner induction field at values $\approx \mu_0$~\hcp; this may imply high dissipation due to both the high quasiparticle density and the free motion of fluxons. These considerations, though may justify the presence of the magnetic hysteresis, do not quantitatively account for the plateau; the latter can be justified only if one assumes that the region of the sample in which the fluxon lattice is in the critical state, after it has been exposed to applied fields higher enough than \hcp, is much smaller than that in which the local induction maintains the value $\mu_0$~\hcp.
\section*{Conclusion}
We have investigated the field-induced variations of the microwave surface resistance of a polycrystalline sample of \mgb, at fixed temperatures. The mw surface resistance has been measured by the cavity-perturbation technique, at increasing and decreasing the DC magnetic field in the range $0 \div 1$~T. The $R_s(H_0)$ curves exhibit a magnetic hysteresis of unconventional shape: in the decreasing-field branch of the $R_s(H_0)$ curve, we have detected an unexpected plateau extending from a certain magnetic-field value down to zero. The hysteresis is detectable up to temperatures close to $T_c$ ($T/T_c\approx 0.95)$; however, for $T\geq 0.77~T_c$ the hysteresis manifests itself only through the presence of the plateau.

It has been shown by different authors that the standard models for the investigation of fluxon dynamics are inadequate to justify the results in \mgb\ in wide ranges of fields and temperatures, due to the unusual vortex structure (double core). Our results confirm this idea; however, in this paper, we have devoted attention to the range of fields at which the superconductivity coming from the $\pi$ band is almost suppressed. In this region ($H_0 \geq H_{c2}^{\pi}$), we expect that the flux lines assume a conventional single-core structure and all the charge carriers coming from the $\pi$ band are quasiparticles. On this hypothesis, we have modified the expression of the complex penetration depth of the mw field, considering that the contribution of the  $\pi$ band to the field-induced energy losses is merely due to the presence of the $\pi$ quasiparticles and that of the $\sigma$ band is due to both the $\sigma$ quasiparticles and the motion of conventional (single core) fluxons. By using the generalized two-band Eliashberg theory, we have calculated the temperature dependence of the $\sigma$-band quasiparticle density, at zero field, which has been successfully used for determining the expected $R_s(H_0, T)$ curves. In the framework of the same theory, we have also calculated the temperature dependence of the upper critical field and compared it with that deduced from the experimental results, in the range of temperatures investigated.

We have quantitatively analyzed the experimental results obtained at temperatures near $T_c$ ($T/T_c \geq 0.77)$, where our experimental apparatus allows reaching DC magnetic fields larger enough than $H_{c2}^{\pi}$. Since the used model is valid only for $H_0 \geq H_{c2}^{\pi}$, by looking at the field range in which  the expected results fit the experimental data, we have determined with good accuracy the temperature dependence of $H_{c2}^{\pi}$ at temperatures near $T_c$, where the standard techniques fail in giving accurate results. Our results show that the applied magnetic field that separates the reversible and irreversible region of the $R_s(H_0)$ curves is just $H_{c2}^{\pi}(T)$.
This interesting result seems to link the magnetic field that suppresses the superconductivity of the $\pi$ band to the pinning efficacy. We have suggested that for $H_0 \geq H_{c2}^{\pi}$ the $R_s(H_0)$ is reversible because the vortex cores are surrounded by both $\sigma$-band condensed fluid and $\pi$-band normal fluid; this reduces the stabilization energy and, consequently, the pinning efficacy. The hysteresis observed for $H_0 \leq H_{c2}^{\pi}$ can be justified assuming that vortex pinning is effective up to temperatures near $T_c$ when charge carriers from both $\pi$ and $\sigma$ bands contribute to the condensed fluid.

The reason for which the hysteretic behaviour of the $R_s(H_0)$ curve at temperatures near $T_c$ manifests itself only through the presence of the plateau, extending from $H_{c2}^{\pi}$ down to zero, is not fully understood and further experimental as well as theoretical investigation is necessary. A possible explanation is that, after the sample has been exposed to DC magnetic fields higher that $H_{c2}^{\pi}$, on reducing $H_0$ down to zero, in a wide inner region of the sample the local induction field maintains values $\approx \mu_0~H_{c2}^{\pi}$ and only in a narrow region near the sample surface, where the local magnetic field is lower than $\mu_0~H_{c2}^{\pi}$, the critical state develops.

\appendix
\section{Calculus of the superfluid density at zero magnetic field}
The generalization of the Eliashberg theory for systems with two bands has already been used with success to study the MgB$_{2}$ system~\cite{carbinicol}. To obtain the gaps, and then the superfluid densities $n_s^{\sigma}(T)$, $n_s^{\pi}(T)$ and the critical temperature, within the $s$-wave two-band Eliashberg model, one has to solve four coupled integral equations for the gaps $\Delta_{i}(i\omega_{n})$ and the renormalization functions $Z_{i}(i\omega_{n})$, where $i=\sigma,\pi$ is a band index and $\omega_{n}$ are the Matsubara frequencies. Including in the equations the non-magnetic- and paramagnetic-impurity scattering rates in the Born approximation, $\Gamma^{ij}_{+}$ and $\Gamma^{ij}_{-}$, one obtains
 \begin{eqnarray}
\omega_{n}Z_{i}(i\omega_{n})&=&\omega_{n}+\pi
T\sum_{m,j}\Lambda_{ij}(i\omega_{n}-i\omega_{m})N^{j}_{Z}(i\omega_{m})+\nonumber\\
& &
+\sum_{j}(\Gamma^{ij}_{+}+\Gamma^{ij}_{-})N^{j}_{Z}(i\omega_{n})\label{eq:EE1}
\end{eqnarray}
\begin{eqnarray}
Z_{i}(i\omega_{n})\Delta_{i}(i\omega_{n})&=&\pi
T\sum_{m,j}[\Lambda_{ij}(i\omega_{n}-i\omega_{m})-\mu^{*}_{ij}(\omega_{c})]\cdot\nonumber\\
& &
\hspace{-2.5cm}\cdot\theta(|\omega_{c}|-\omega_{m})N^{j}_{\Delta}(i\omega_{m})+\sum_{j}%
(\Gamma^{ij}_{+}-\Gamma^{ij}_{-})N^{j}_{\Delta}(i\omega_{n})
\label{eq:EE2}
\end{eqnarray}
where $\theta$ is the Heaviside function, $\omega_{c}$ is a
cut-off energy, and
\begin{equation*}
    \Lambda_{ij}(i\omega_{n}-i\omega_{m})=\int_{0}^{+\infty}d\omega
\alpha^{2}_{ij}F(\omega)/[(\omega_{n}-\omega_{m})^{2}+\omega^{2}]\,,
\end{equation*}
\begin{equation*}
    N^{j}_{\Delta}(i\omega_{m})=\Delta_{j}(i\omega_{m})/
{\sqrt{\omega^{2}_{m}+\Delta^{2}_{j}(i\omega_{m})}}\,,
\end{equation*}
\begin{equation*}
    N^{j}_{Z}(i\omega_{m})=\omega_{m}/
{\sqrt{\omega^{2}_{m}+\Delta^{2}_{j}(i\omega_{m})}}\,.
\end{equation*}
The solution of the Eliashberg equations requires as input:\\i) the four electron-phonon spectral functions $\alpha^{2}_{ij}(\omega)F(\omega)$;\\
ii) the four  elements of the Coulomb pseudopotential matrix $\mu^{*}(\omega_{c})$;\\iii) the four non-magnetic-impurity scattering rates $\Gamma_+^{ij}$;\\iv) the four paramagnetic-impurity scattering rates $\Gamma_{-}^{ij}$.\\
In our calculus is always $\Gamma_+^{ij}=\Gamma_-^{ij}=0$. The four spectral functions $\alpha^{2}_{ij}(\omega)F(\omega)$, which were calculated for pure
MgB$_{2}$ in Ref.~\cite{GolubovA2F}, have the following electron-phonon coupling constant:
$\lambda_{\sigma\sigma}$=1.017, $\lambda_{\pi\pi}$=0.448, $\lambda_{\sigma\pi}$=0.213 and
$\lambda_{\pi\sigma}$=0.156 \cite{GolubovA2F}.

As far as the Coulomb pseudopotential is concerned, we use the
expression calculated for pure MgB$_2$~\cite{DolgovCoulomb}
\begin{eqnarray}
\hspace{-5mm}\mu^{*}\!\!&\!=~\!&\!\! \left| \begin{array}{cc}%
\mu^{*}_{\sigma \sigma} & \mu^{*}_{\sigma \pi}\\
\mu^{*}_{\pi \sigma} & \mu^{*}_{\pi \pi}
\end{array} \right| =  \nonumber \\
\!\!&\!=~\!&\!\! \mu(\omega_{c})N_{N}^{tot}(E_{F})
\left| \begin{array}{cc}%
\frac{2.23}{N_{N}^{\sigma}(E_{F})} &
\frac{1}{N_{N}^{\sigma}(E_{F})}\\ & \\
\frac{1}{N_{N}^{\pi}(E_{F})} &
\frac{2.48}{N_{N}^{\pi}(E_{F})}
\end{array} \right| \label{eq:mu}
\end{eqnarray}
where $\mu(\omega_{c})$ is a free parameter, $N_{N}^{i}(E_{F})$ is the normal density of states at the Fermi level of the i-band and
$N_{N}^{tot}(E_{F})=N_{N}^{\sigma}(E_{F})+N_{N}^{\pi}(E_{F})$.\\
For obtaining the experimental critical temperature we fix $\mu(\omega_{c})=0.03153$ with cut-off energy $\omega_{c}=450$ meV and maximum energy 500 meV. In all our calculations we use
$N_{N}^{\sigma}(E_{F})=0.30$ states/(cell eV) and $N_{N}^{\pi}(E_{F})=0.41$ states/(cell eV)~\cite{golubov}.

\section{Calculus of the temperature dependence of the upper critical field}
In order to calculate the upper critical field we have used the linearized gap equations under magnetic field, for a superconductor in the clean limit (negligible impurity scattering) \cite{Sud}
 \begin{eqnarray}
\omega_{n}Z_{i}(i\omega_{n})&=&\omega_{n}+\pi
T\sum_{m,j}\Lambda_{ij}(i\omega_{n}-i\omega_{m})sign(\omega_{m})
\end{eqnarray}
\begin{eqnarray}
Z_{i}(i\omega_{n})\Delta_{i}(i\omega_{n})&=&\pi
T\sum_{m,j}[\Lambda_{ij}(i\omega_{n}-i\omega_{m})-\mu^{*}_{ij}(\omega_{c})]\cdot\nonumber\\
& &
\hspace{-1.5cm}\cdot\theta(|\omega_{c}|-\omega_{m})\chi_{j}(i\omega_{m})Z_{j}(i\omega_{m})\Delta_{j}(i\omega_{m})
\end{eqnarray}
\begin{eqnarray}
\chi_{j}(i\omega_{m})&=&\frac{2}{\sqrt{\beta_j}}\int^{+\infty}_{0}dq\exp(-q^{2})\cdot\nonumber\\
& &
\hspace{-1.5cm}\cdot tan^{-1}\frac{q\sqrt{\beta_{j}}}{|\omega_{m}Z_{j}(i\omega_{m})|+i\mu_{B}H_{c2}sign(\omega_{m})}\nonumber\\
\end{eqnarray}
with $\beta_{j}=\pi (v^j_{F})^{2} H_{c2}/(2\Phi_{0})$; $v^j_{F}$ is the Fermi velocity of band j and $H_{c2}$ is the upper critical field.\\
In these equations, the bare Fermi velocities are the input parameters for calculating the upper critical field as a function of temperature. To obtain the theoretical curve reported in Figure~\ref{Hc2(T)}, we have used $v^{\sigma}_{F ab}=4.4\times 10^{5}$~m/s and $v^{\pi}_{F ab}=20\times 10^{5}$~m/s.

\ack
The authors are very glad to thank P. Manfrinetti for having kindly supplied the \mgb\ sample; I. Veshchunov and Yu. A. Nefyodov for having done the SEM micrograph of the sample surface; G. Lapis and G. Napoli for technical assistance.


\section*{References}
\begin{thebibliography}{99}

\bibitem{golo} Golosovsky M, Tsindlekht M and Davidov D 1996 \textit{Supercond. Sci. Technol.}
\textbf{9} 1 and Refs. therein.

\bibitem{GOLOSOVSKY} Golosovsky M, Tsindlekht M, Chayet H and Davidov D 1994 \textit{Phys. Rev.}
B {\bf 50} 470.

\bibitem{noistatocritico} Bonura M, Di Gennaro E, Agliolo Gallitto A and
Li Vigni M 2006 \textit{Eur. Phys. J.} B {\bf 52} 459.

\bibitem{noiisteresi} Bonura M,  Agliolo Gallitto A and Li Vigni M 2006
\textit{Eur. Phys. J.} B {\bf 53} 315.

\bibitem{CC} Coffey M W and Clem J R 1991 \textit{Phys. Rev. Lett.} {\bf67} 386;
Coffey M W and Clem J R 1992 \textit{Phys. Rev.} B {\bf 45} 9872; Coffey M W and Clem J R 1992 \textit{Phys. Rev.} B {\bf 45} 10527.

\bibitem{BRANDT} Brandt E H 1991 \textit{Phys. Rev. Lett.} {\bf 67} 2219.

\bibitem{dulcicvecchio} Dul$\check{\mathrm{c}}$i$\acute{\mathrm{c}}$ A and
Po$\check{\mathrm{z}}$ek M 1993 \textit{Physica} C \textbf{218} 449.

\bibitem{shibata} Shibata A, Matsumoto M, Izawa K, Matsuda Y, Lee S
and Tajima S 2003 \textit{Phys. Rev.} B {\bf 68} 060501(R).

\bibitem{nova} Agliolo Gallitto A, Bonsignore G, Fricano S, Guccione M and Li Vigni M 2005
\emph{Topics in Superconductivity Research}, ed B P Martins
(New York: Nova Science Publishers) pags. 125-143.

\bibitem{dulcic} Dul$\check{\mathrm{c}}$i$\acute{\mathrm{c}}$ A, Paar D, Po$\check{\mathrm{z}}$ek M, Williams V M, Kr$\ddot{\mathrm{a}}$mer S, Jung C U, Min-Seok Park and Sung-Ik Lee 2002 \textit{Phys. Rev.} B {\bf 66} 014505.

\bibitem{noi-irr} Bonura M, Agliolo Gallitto A, Li Vigni M, Ferdeghini C
and Tarantini C 2008 \textit{Eur. Phys. J.} B \textbf{63} 165.

\bibitem{isteresiMgB2} Agliolo Gallitto A, Bonura M, Fricano S, Li Vigni M and
Giunchi G 2003 \textit{Physica} C {\bf 404} 171.

\bibitem{EUCAS2007} Agliolo Gallitto A, Bonura M and Li Vigni M 2008 \textit{J. Phys.: Conf. Ser.} {\bf 97} 012207.

\bibitem{eskil} Eskildsen M R, Kugler M, Tanaka S, Jun J, Kazakov S M,
Karpinski J and Fisher $\o$ 2002 \textit{Phys. Rev. Lett.} {\bf 89} 187003.

\bibitem{nakai} Nakai N, Ichioka M and Machida K 2002 \textit{J. Phys. Soc. Jpn.} \textbf{71} 23.

\bibitem{koshelev} Golubov A A and Koshelev A E 2003 \textit{Phys. Rev. Lett.} \textbf{90}
177002.

\bibitem{bouquet-H} Bouquet F, Wang Y, Sheikin I, Plackovski T and Junod A 2002
\textit{Phys. Rev. Lett.} {\bf 89} 257001.

\bibitem{Daghero2} Daghero D, Gonnelli R S, Ummarino G A, Stephanov V A, Jun J,
Kazakov S M and Karpinski J 2003 \textit{Physica} C {\bf 355} 255.

\bibitem{Samuely} Samuely P, Szab$\acute{\mathrm{o}}$ P, Ka$\check{\mathrm{c}}$mar$\check{\mathrm{c}}\acute{\mathrm{i}}$k J, Klein T and Jansen A G M 2003 \textit{Physica} C {\bf 355} 244.

\bibitem{diffusivita} Bogoslavsky Y, Miyoshi Y, Perkins G K, Kaplin A D, Cohen L F, Progrebnyakov A V and Xi X X 2005 \textit{Phys. Rev.} B {\bf 72} 224506.

\bibitem{Sarti} Sarti S, Amabile C, Silva E, Giura M, Fastampa R, Ferdeghini C, Ferrando V and Tarantini C 2005 \textit{Phys. Rev.} B \textbf{72} 024542.

\bibitem{palenzona} Palenzona A, Manfrinetti P and Braccini V 2001 INFM patent n. TO2001A001098.

\bibitem{Ji} Ji L, Rzchowski M S, Anand N and Tinkham M 1993 \textit{Phys. Rev.} B {\bf 47} 470.

\bibitem{sridhar} Willemsen B A, Derov J S and Sridhar S 1997 \textit{Phys. Rev.} B {\bf 56}
11989.

\bibitem{noiBKBO} Fricano S, Bonura M, Agliolo Gallitto A, Li Vigni A, Klinkova L A and Barkovskii N V 2004 \textit{Eur. Phys. J.} B {\bf 41} 313.

\bibitem{TALVA} Owliaei J, Shridar S and Talvacchio J 1992 \textit{Phys. Rev. Lett.} {\bf 69} 3366.

\bibitem{noi-depin} Bonura M, Agliolo Gallitto A, Li Vigni M and Martinelli A 2008 \textit{Physica} C \textbf{468} 2372.

\bibitem{Mazin:107002} Mazin I I, Andersen O K, Jepsen O, Dolgov O V, Kortus J, Golubov A A,
Kuz'menko A B and Van der Marel D 2002 \textit{Phys. Rev. Lett.} \textbf{89} 107002.

\bibitem{gittle} Gittleman J I and Rosenblum B 1996 \textit{Phys. Rev. Lett.} {\bf 16} 734.

\bibitem{silva} Sarti S, Amabile C, Fastampa R, Giura M, Pompeo N and Silva E 2007 \textit{J. Supercond.} \textbf{20} 51.

\bibitem{golubov} Golubov A A and Koshelev A E 2003 \textit{Phys. Rev.} B \textbf{68}104503.

\bibitem{cubit} Cubitt R, Levett S, Bud'ko S L, Anderson N E and Canfield P C 2003 \textit{Phys. Rev. Lett.} {\bf 90} 157002; Cubitt R, Eskildsen M R, Dewhurst C D, Jun J, Kazakov S M and Karpinski J 2003 \textit{Phys. Rev. Lett.} {\bf 91} 047002.

\bibitem{Lyard} Lyard L, Szab\'{o} P, Klein T, Markus J, Marcenat C, Kim K H, Kang B W, Lee H S and Lee S I 2004 \textit{Phys. Rev. Lett.} \textbf{92} 57001.

\bibitem{Li} Li S L, Wen H H, Zhao Z W, Ni Y M, Ren Z A, Che G C, Yang H P, Liu Z Y and Zhao Z X 2001 \textit{Phys. Rev.} B \textbf{64} 094522.

\bibitem{sharoni} Sharoni A, Felner I and Millo O 2001 \textit{Phys. Rev.} B \textbf{63} 220508(R).

\bibitem{kortus} Kortus J, Mazin I I, Belashchenko K D, Antropov V P and Boyer L L 2001 \textit{Phys. Rev. Lett.} \textbf{86} 4656.

\bibitem{Brinkman} Brinkman A, Golubov A A, Rogalla H, Dolgov O V, Kortus J, Kong Y, Jepsen O and Andersen O K 2002 \textit{Phys. Rev.} B \textbf{65} 180517(R).

\bibitem{carbinicol} Nicol E J and Carbotte J P 2005 \textit{Phys. Rev.} B \textbf{71} 054501.

\bibitem{Sud} Suderow H, Tissen V G, Brison J P, Martínez J L, Vieira S, Lejay P, Lee S and Tajima S 2004 \textit{Phys. Rev.} B \textbf{70} 134518; Ummarino G A 2005 \textit{Physica} C \textbf{423} 96.

\bibitem{GolubovA2F} Golubov A A, Kortus J, Dolgov O V, Jepsen O, Kong Y, Andersen O K, Gibson B J, Ahn K and Kremer R K 2002 \textit{J. Phys.: Condens. Matter} \textbf{14} 1353.

\bibitem{DolgovCoulomb} Dolgov O V 2003 Talk at the $6^{th}$ European Conference on Applied Superconductivity (EUCAS), Sorrento, Italy, September 14-18, 2003; Mazin I I, Andersen O K, Jepsen O, Golubov A A, Dolgov O V and Kortus J 2004 \textit{Phys. Rev.} B \textbf{69} 056501.
\end{thebibliography}
\end{document}